\documentclass[journal]{IEEEtran}
\pdfoutput=1
\usepackage{graphicx}
\usepackage{cite}
\usepackage{amsmath}
\usepackage[numbers]{natbib}
\usepackage{color}
\begin{document}
\markboth{4EPo2E-07}%
{4EPo2E-07}
\title{Heat transfer coefficient saturation in superconducting Nb tunnel junctions contacted to a NbTiN circuit and an Au energy relaxation layer}
\author{Stefan~Selig, Marc~Peter~Westig, Karl~Jacobs, Michael~Schultz and Netty~Honingh
\thanks{
S.~Selig, K.~Jacobs, M.~Schultz and N.~Honingh are with the I.~Physikalisches Institut, 
Universit{\"a}t zu K{\"o}ln, 50937 K{\"o}ln, Germany, 
e-mail: (selig@ph1.uni-koeln.de).}
\thanks{M.P.~Westig is with the Service de Physique de l'Etat Condens\'{e} (CNRS URA 2464), IRAMIS, CEA Saclay, 
91191 Gif-sur-Yvette, France.}
}
\IEEEpubid{10.1109/TASC.2014.2378054~\copyright~2014 IEEE}
\maketitle
\begin{abstract}
In this paper we present the experimental realization of a Nb tunnel junction connected to a 
high-gap superconducting NbTiN embedding circuit. We investigate relaxation of nonequilibrium quasiparticles
in a small volume Au layer between the Nb tunnel junction and the NbTiN circuit. We find a saturation in the 
effective heat-transfer coefficient consistent with a simple theoretical model. This saturation is 
determined by the thickness of the Au layer. Our findings are important for the
design of the ideal Au energy relaxation layer for practical SIS heterodyne mixers and we suggest two geometries, 
one, using a circular Au layer and, two, using a half-circular Au layer. Our work is concluded with an outlook of our future
experiments.
\end{abstract}
\IEEEpeerreviewmaketitle
\section{\label{sec:01}Introduction}
In radio astronomy, heterodyne instruments are used to detect weak radiation from an astronomical source with a 
very high spectral resolution \cite{deGraauw2010}. 
The central element of a heterodyne instrument is a mixer which has a strong nonlinear current-voltage (I-V) characteristic. 
In this article we will focus on superconductor-insulator-superconductor (SIS) tunnel junctions since they show a 
highly nonlinear I-V characteristic due to the singularity in the superconducting 
quasiparticle density-of-states and, therefore, perfectly fulfill the needs of a high-sensitivity astronomical heterodyne receiver. 
For a more complete review of other mixers, like the hot-electron bolometer mixer, we refer to \cite{zmuidzinas2004}.

The frequency mixing limit (i.e.~the highest detectable frequency) in a SIS mixer is set by the superconducting 
pair-breaking energy $2\Delta$. In this paper we focus on a symmetric SIS mixer which uses the same 
superconducting material for both junction electrodes. In such a device the frequency mixing limit is 
$\nu_{max} = 4\Delta/h$ where $h$ is the Planck constant. 
Existing receivers almost enirely use Nb-Al-AlOx-Nb junctions, because they can be reliably 
fabricated with high quality, i.e.~with very low subgap/leakage currents. 
Moreover, these devices already provide a reasonably high frequency mixing limit of theoretically 1.4~THz making them 
interesting for applications in radio-astronomy. In order to achieve the fundamental sensitivity limit of a 
SIS mixer \cite{tucker1985}, it is crucial that the tunnel junction is embedded in a high-Q tuning circuit that enables a 
near 100~\% coupling to the sky signal. For this reason optimum embedding circuits are 
made of superconducting transmission lines. However, here 
lossless signal transport is provided only up to the superconducting 
pair-breaking frequency $2\Delta/h$. Up to the pair-breaking frequency of niobium of approximately 
700~GHz, the embedding circuit is made of the same material like the Nb junction electrodes. 
Near quantum-limited sensitivity for such a device has been 
demonstrated over a broad range of frequencies \cite{zmuidzinas2004}.
For larger frequencies for which $h\nu > 2\Delta$, quasiparticle 
excitations in the embedding circuit are created due to the breaking of Cooper pairs and large ohmic 
losses are observed which would significantly decrease the sensitivity of a mixer device. On the other hand, 
heterodyne mixing above the pair-breaking frequency is not intrinsically limited by the superconducting electrode
material itself, although here also quasiparticle excitations are created, absorbing a part of the detection signal.
\IEEEpubidadjcol

Consequently, in order to exploit the full frequency mixing range of a high-sensitivity Nb SIS mixer, 
an embedding circuit made of superconducting transmission lines with a higher superconducting pair-breaking 
energy than the junction electrodes are needed. E.g.~in \cite{Jackson2006, Kroug2009} experiments are 
reported where the material NbTiN was used as superconducting embedding material contacted to only 
one electrode of the SIS mixer whereas the other side of the SIS junction was contacted to the good 
normal-conducting metal Al. Losses in these devices due to the normal-conducting Al layer can be
prevented by replacing it against NbTiN \cite{Leone2000,Jackson2005} resulting in Nb SIS junctions
contacted to a full superconducting NbTiN embedding circuit. An important conclusion of this earlier work was, that it turned 
out to be important to avoid heating effects of the electron gas of the Nb junction due to blocked quasiparticle outdiffusion 
at the Andreev trap formed between the high-gap material NbTiN and the lower-gap material Nb. 

We experimentally found the relaxation of these nonequilibrium quasiparticles in 
an added small normal-metal layer of Au between the 
Nb junction and the NbTiN embedding circuit \cite{westig2013}. In this proof-of-principle study, the Au 
layer had a circular shape with a diameter of several $\mu$m, centered on the SIS junction. This is not necessarily the
best geometry for a practical heterodyne mixer because the large overlap of the Au layer and the NbTiN layer which is one of the electrodes of the radio frequency (RF) tuning circuit of the SIS junction, 
causes RF signal losses, that depend on the length of the overlap. However, we have found in a different experiment, that the 
relaxation of nonequilibrium quasiparticles does not depend on the particular geometry of the Au layer as long as 
the volume of the Au layer is sufficient for the relaxation \cite{Selig2014}. 
Possible remaining influences of the geometry on the relaxation process could not be identified so far.

\begin{figure}[tb]
\centering
\includegraphics[width=0.85\columnwidth]{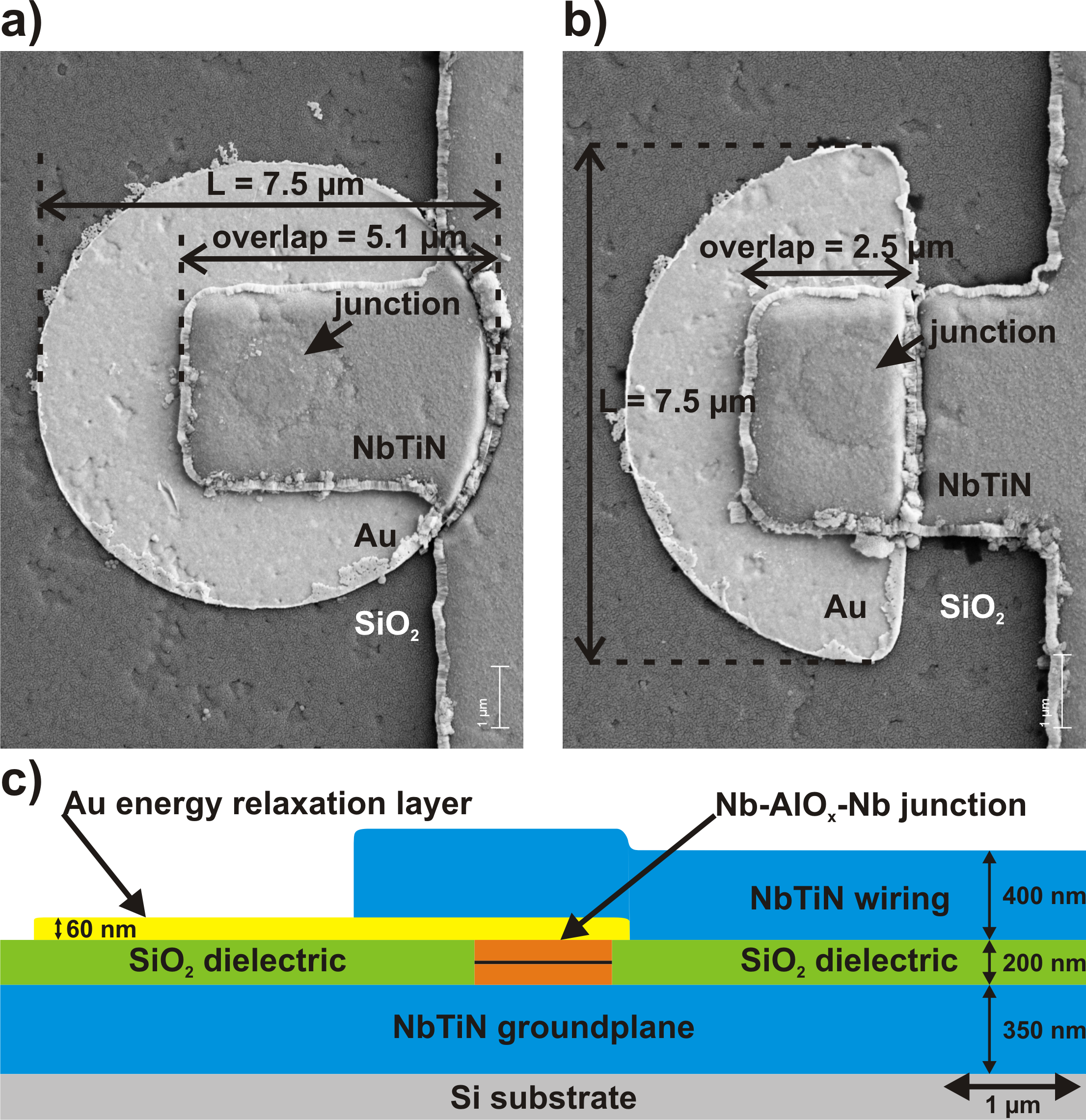}  
\caption{Scanning electron microscopy pictures showing Nb SIS tunnel junctions contacted to a 
NbTiN embedding circuit. (a) shows a circular Au energy relaxation layer between the Nb SIS tunnel junction and the NbTiN layer and (b) a half-circular Au energy relaxation layer. (c) shows a schematic cross section of a device with half-circular Au layer.}
\label{fig:01}
\end{figure}

With these findings one would in principle favor, first, an Au layer shape 
with a small overlap with the NbTiN embedding circuit and, second, an Au layer with a
minimum volume. A shape with these properties is e.g.~a half circular shape (compare with Fig.~{\ref{fig:01}}).

In this article we present our ongoing work and focus in particular on a thorough systematic study of the 
relaxation of nonequilibrium quasiparticles in various circular and half-circular shaped Au layers, added between the
Nb junction and the NbTiN embedding circuit. 
Earlier work \cite{Selig2014} showed  a linear dependence between the volume of the Au layer and the heat-transfer coefficient alpha. On further increasing the volume of the Au-layer while keeping its thickness fixed to $d_{Au} = 60$~nm, we find in dc-measurements of the I-V characteristic a saturation in the nonequilibrium quasiparticle relaxation.
This suggests an effect of the geometry on the relaxation process and we supplement our earlier findings 
by this new result. 

We reproduced our experimental 
results with a simple theoretical model of our device. With this model we can also consistently describe the difference in the heating of the Nb electron gas of a device which we immersed in liquid helium (LHe) and
which was characterized in a vacuum dewar. With this stage of consistency in our description, 
this allows us to determine the optimal Nb mixer device consisting of a full superconducting NbTiN embedding 
circuit and an Au energy relaxation layer. 

In Sec.~\ref{sec:02} we present briefly our theoretical model and our method to calculate the heat-transfer
coefficient $\alpha$ as a function of Au layer volume and geometry.
In Sec.~\ref{sec:03} we describe our experiments and present the measurement results. 
Section~\ref{sec:04} presents our conclusion.
\section{\label{sec:02}Energy Relaxation and Heat balance equation}
Considering only lowest order tunneling processes, the quasiparticle current through a symmetric SIS junction is
given by
\begin{equation}
\label{eq:01}
\begin{split}
I_{qp}(V) = \frac{1}{eR_{N}} \int_{-\infty}^{+\infty}dE~&N_{S}(E,\Delta)N_{S}(E+eV,\Delta)\\
&\cdot\left[f(E,T) - f(E+eV,T)\right]~,
\end{split}
\end{equation}
with $V$ the bias voltage applied to the junction, $f(E,T)$ the Fermi-Dirac distribution function 
of quasiparticles of energy $E$ having an effective electron temperature $T = T_{e}$ and the quantities $N_{S}$ are the 
normalized BCS density of states of the electrodes \cite{Bardeen1957}. The normal-state resistance of the tunnel junction, $R_{N}$, 
is a measure for the transmissivity of the tunnel barrier. In order to provide a large RF bandwidth of a SIS mixer, one
prefers low $R_{N}A$ products, or high current densities, where $A$ is the area of the tunnel junction \cite{tucker1985}.
By using Eq.~(\ref{eq:01}) one assumes that the tunnel current does not disturb the physical properties of the two junction 
electrodes. Hence, one uses the Fermi-Dirac distribution at the electron temperature $T_{e}$ and the value for $\Delta$ at the
same temperature. In equilibrium, $T_{e}$ is equal to the phonon bath temperature $T_{ph}$.

For a quasiparticle nonequilibrium distribution due to dissipated power in the SIS junction, $T_{e}>T_{ph}$. 
This elevated $T_{e}$ directly influences the Fermi-Dirac distribution and consequently the superconducting
energy gap $\Delta$ through the self-consistency relation \cite{Bardeen1957}
\begin{equation}
\label{eq:02}
\frac{1}{N(E_{F})\mathcal{V}} = \int_{0}^{k_{B}\theta_{D}}d\epsilon~
\frac{1-2f\left[(\epsilon^{2} + \Delta(T_{e})^2)^{1/2}\right]}{(\epsilon^{2} + \Delta(T_{e})^2)^{1/2}}~,
\end{equation}
where $N(E_{F})$ is the single-spin density of states at the Fermi 
energy in the normal state, $k_{B}$ is the Boltzmann constant, 
$\theta_{D}$ is the Debye temperature of the material, $\mathcal{V}$ is the average attractive 
interaction potential of superconductivity describing phonon exchange between 
electrons \citep{Bardeen1957} and $\epsilon$ is the independent 
quasiparticle energy measured relative to the Fermi energy. The temperature $T_{e}$ depends on the power generated by the 
tunneling process. The power rapidly rises for bias voltages $V_g = 2\Delta/e$. In case that this increases $T_{e}$, 
Eq.~(\ref{eq:02}) implies that the relevant energy gap for the tunneling decreases leading to the possibility of 
back-bending \cite{Leone2000, westig2013, Selig2014} (the slope of $I_{qp}$ becomes negative around $V_{g}$). 
Therefore, the slope of $I_{qp}$ directly probes $T_{e}$ over the reduction in $\Delta(T_{e})$.
In order to deal with non-idealities in the sharpness of the I-V curve compared to the ideal situation described by Eq.~(\ref{eq:01}), 
a phenomenological broadening parameter $\Gamma$ is introduced into the expression for $N_{S}$ 
allowing to estimate the smearing of $I_{qp}$ around $V_{g}$ due to quasiparticle energy 
broadening \cite{dynes1978, westig2013}, e.g.~caused by an external magnetic field which we use to suppress the Josephson
effect during our measurements. This magnetic field also closes the small mini-gap in the Au-layer that developes due to the proximity effect.

The actual value of $T_{e}$ depends on the energy relaxation channels in the device. The injection of hot electrons by the tunnel barrier
into the junction electrodes is followed by the fast electron-electron relaxation with characteristic time $\tau_{e-e}$. 
This process is usually fast enough so that one can assume a homogeneous effective electron temperature $T_{e}$. In a closed volume 
where diffusion processes play no role, $T_{e}$ is relaxed via the slower electron-phonon relaxation 
with characteristic time $\tau_{e-ph}$. In an open volume, like it is the case for our small Au layer, diffusion processes play a role and 
provide an additional relaxation channel for quasiparticles.  
This balancing of $T_{e}$ via the aforementioned processes is described by the following general
heat-balance equation \cite{skocpol1974}
\begin{equation}
\label{eq:03}
-\kappa \left[r^2\frac{d^2T_{e}(r)}{dr^2} + r\frac{dT_{e}(r)}{dr}\right] + 
\frac{Y}{d_{Au}} r^2 \left[T_{e}(r)-T_{ph}\right] = 0~.	
\end{equation}
Its solution describes the radial temperature distribution around the SIS junction in the 
Au layer shown in Fig.~\ref{fig:01}. 
The thermal conductivity is denoted by $\kappa$, $r$ is the radial
position, measured from the center of the SIS junction, $Y$ is the heat-transfer coefficient (measured in $\mathrm{W/m^2 K}$) to the phonon system, $T_{e}(r)$ is the spatially dependent electron 
temperature in the Au layer, $T_{ph}$ is the bath (phonon) temperature as before and 
$d_{Au}$ is the thickness of the Au layer. 
We distinguish between the metallic phonon system, 
where $Y = Y_{e-ph}$ with $Y_{e-ph}$ being the metallic
electron-phonon heat-transfer coefficient and the heat transfer to the phonon bath system (substrate of LHe). In this case is $Y = Y_{K}$ with $Y_{K}$ being the Kapitza conductance between the metal and the substrate or between the metal and the LHe. For our interpretation it turned out that the dominating relaxation channel 
(i.e.~the slowest relaxation process) is given by the Kapitza resistance, so we set $Y = Y_{K}$ for further calculations.

Furthermore, for the interpretation of our experiments, we use a second linearized heat balance equation
\begin{equation}
\label{eq:04}
P = \alpha \left(T_{e} - T_{ph}\right)~.
\end{equation}
Here, $\alpha$ is an effective heat-transfer coefficient. Comparing Eq.~(\ref{eq:04}) and Eq.~(\ref{eq:03}), for a 
distributed layer of Au where quasiparticle relaxation via diffusion and electron-phonon scattering occurs 
simultaneously, $\alpha$ contains both relaxation contributions. This is in contrast to the definition of $\alpha$ we used
in our earlier work \cite{westig2013} where $\alpha$ was used to describe relaxation only via heat transfer to the phonon system. 
By calculating $T_{e}$ from Eq.~(\ref{eq:03}), the effective $\alpha$ can be determined by using Eq.~(\ref{eq:04}).
This provides a way for direct comparison with experiments since we determine the effective $\alpha$ experimentally 
over a measurement of $I_{qp}$ at voltages around $V_{g}$ from which we obtain $\Delta(T_{e})$. 
Finally, by inverting Eq.~(\ref{eq:02}) we obtain $T_{e}$ and similarly $\alpha$. 
More details are described in \cite{westig2013, Selig2014}. 

In order to describe our device geometries, i.e.~in particular the circular and half-circular shaped Au layers,
Eq.~(\ref{eq:03}) has to be supplemented by sufficient boundary conditions. These read
 \begin{equation}
\label{eq:05}
-\kappa \frac{dT_{e}(r)}{dr} = \frac{IV}{(2)\cdot\sqrt{A\pi}d_{Au}},
~\quad\mathrm{for}~r\rightarrow r_{j}=\sqrt{\frac{A}{\pi}}~,
\end{equation}
and 
\begin{equation}
\label{eq:06}
\frac{dT_{e}(r)}{dr} = 0,~\quad\mathrm{for}~r\rightarrow r_{C}.
\end{equation}
\begin{figure}[tb]
\centering
\includegraphics[width = 0.9\columnwidth]{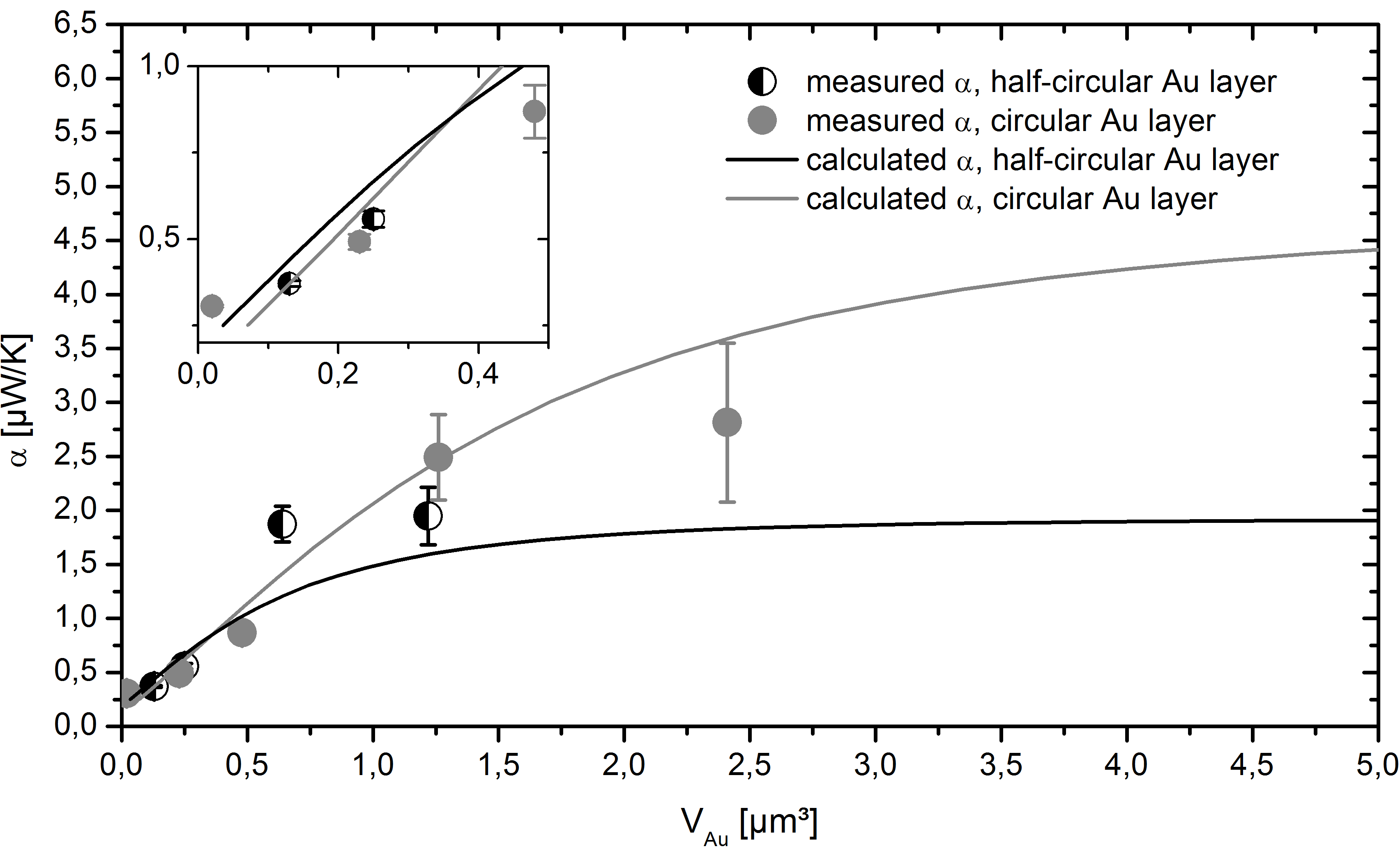}  
\caption{Saturation in the effective heat-transfer coefficient $\alpha$. $V$ is the volume of the Au layer
with constant thickness $d_{Au}$ and the volume is increased by only increasing $r_{C}$. Data points are experimental results averaged over a large number of fabricated devices.}
\label{fig:02}
\end{figure}
The boundary condition in Eq.~(\ref{eq:05}) describes the power dissipation in the SIS junction of radius $r_{j}$ 
and subsequent inflow of heat into a circular shaped Au layer when the factor of "2" is used and otherwise 
inflow of heat into the half-circular shaped Au layer. Equation~(\ref{eq:06}) ensures continuity of the solution at the 
boundary of the Au layer with radius $r_{C} = L/2$ (compare with Fig.~{\ref{fig:01}}). Furthermore, $d_{Au}$ is the
thickness of the Au layer and $A$ is the area of the SIS junction which is 1~$\mu\mathrm{m}^2$ in our experiments.
\section{\label{sec:03}Experiment and Measurement Results}
We have fabricated Nb-Al-$\mathrm{AlO_{x}}$-Nb SIS tunnel junctions embedded in a NbTiN-SiO$_2$-NbTiN 
circuit with an Au energy relaxation layer between the junction and the top NbTiN layer. We 
varied the size of the Au layer for a fixed thickness of $d_{Au} = 60$~nm and realized circular and half-circular shapes like 
shown in Fig. \ref{fig:01}. For this paper we performed two measurements with these devices, briefly described below. 
In measurement \#1 we obtained I-V curves of a large number of devices in a dipstick setup, where the devices were immersed 
in LHe. A weak magnetic field was used to suppress the Josephson effect in order to obtain only the quasiparticle current $I_{qp}$. 
We obtained the electron temperature $T_{e}$ as a function of dissipated power $P$ from our
experimental data and extracted $\alpha$ via Eq.~(\ref{eq:04}). In 
Fig.~\ref{fig:02} we have plotted the measured $\alpha$ as a function of Au volume. In order to compare our results with theory
we use the framework described in Sec.~\ref{sec:02}. We solved Eq.~(\ref{eq:03}) while keeping the Au 
thickness $d_{Au} = 60$~nm fixed and varied the radius $r_{C}$ of the Au layer in order to increase its volume $V$. 
This volume is shown on the x-axis of Fig.~\ref{fig:02}. 
The effective heat-transfer coefficient $\alpha$ is obtained by evaluating $T_{e}$ in Eq.~(\ref{eq:03}) using $\kappa = 7~\mathrm{W}/\mathrm{mK}$,
$Y = Y_{K} = 8.9~\cdot10^{4}~\mathrm{W}/\mathrm{m^{2}K}$ \cite{johnson1963} as a function of dissipated power $P$. The result for $T_{e}$ is substituted in Eq.~(\ref{eq:04}) and after this
solved for $\alpha$. 
In our earlier work \cite{westig2013} we included a conservative correction term of 
$-0.15~\mathrm{K}/\mu\mathrm{W}$ to the calculated $\alpha$ in order to account for quasiparticle 
energy broadening $\Gamma$. In this work we increased it to $-0.25~\mathrm{K}/\mu\mathrm{W}$ which leads to a better agreement with our measured data. This modification is reasonable because small changes in e.g.~the magnetic field used to suppress the Josephson
effect in our devices might very well lead to this slight change in the quasiparticle energy broadening correction.
Furthermore, we added an offset to the calculated $\alpha$ equal to
the heat-transfer coefficient of the Nb junction alone. 
Our theoretical results are shown as lines in Fig.~\ref{fig:02} and are in good agreement with the measured data shown as points.
We could also reproduce this result by increasing the values for $\kappa$ and $Y_{K}$ both at the same time by 30~\% but keeping the energy broadening correction term at $-0.15~\mathrm{K}/\mu\mathrm{W}$. This could be attributed to possible changes in our Au film parameters which were not remeasured for this work. Because our fabrication process is in general very stable the modification of the energy broadening correction term is more reasonable.
For our measured data as well as for the theoretical prediction we find a saturation in $\alpha$ when we increase
the volume by increasing only $r_{C}$. On the other hand, when the volume is increased by increasing $d_{Au}$ and possibly
also $r_{C}$, $\alpha$ increases as well. This result is crucially important for the design of Au energy relaxation layers since we find 
that increasing $\alpha$ is limited by the value $d_{Au}$.

\begin{figure}
\centering
\includegraphics[width=0.9\columnwidth]{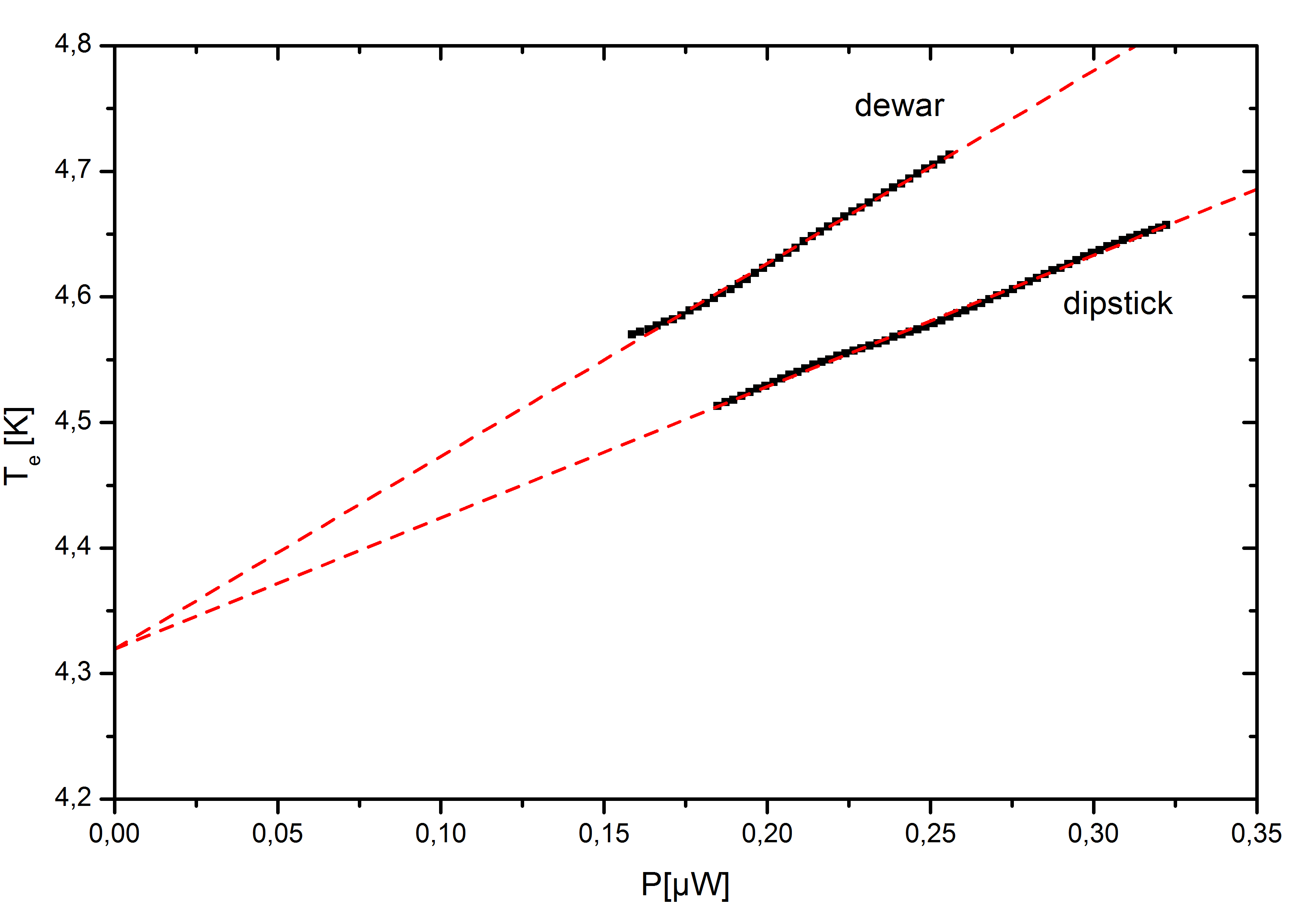}  
\caption{Effective electron temperature $T_{e}$ as a function of dissipated power $P$. The plot shows two 
measurements, first, with the device fixed in a vacuum dewar and, second, with the device fixed in a dipstick and 
directly immersed in LHe.}
\label{fig:03}
\end{figure}

In measurement \#2 we measured the I-V characteristic of a device mounted in a vacuum dewar, like it would be used in 
an astronomical receiver.
In this case the significantly smaller Kapitza conductance between Au and $\mathrm{SiO_{2}}$ 
results in a stronger heating effect than one would obtain for the same measurement with the device immersed in a
LHe vessel where the Kapitza conductance is three times larger (values for the Kapitza resistances we used can be found in
\cite{holt1966, johnson1963}). In Fig.~\ref{fig:03} we show measurements of the
effective electron temperature $T_{e}$ as a function of dissipated power $P$ in a device measured in the 
vacuum dewar and when it is directly immersed in LHe. We adjusted $T_{ph}$ to approximately $4.3~\mathrm{K}$ in the LHe 
vessel for better comparison since this was the bath temperature in the vacuum dewar.
The dimensions of the circular Au layer in the device are $r_{C} = 2.5~\mu$m 
and $d_{Au} = 60~\mathrm{nm}$. Within the framework presented in Sec.~\ref{sec:02}, we can consistently describe 
both measurements for the effective electron temperature $T_{e}$ against dissipated power $P$. 
In order to relax $T_{e}$ towards $T_{ph}$ and when the device is operated in the vacuum dewar, the volume of the Au energy
relaxation layer has to be significantly increased due to the much smaller Kapitza conductance compared 
to the larger Kapitza conductance in LHe. Using our theoretical model, we find that the Au cap should then 
have the dimensions $r_{C} = 7~\mu$m and $d_{Au} = 120~\mathrm{nm}$ ($V = 19~\mu\mathrm{m}^3$) for which our model yields 
an effective heat-transfer coefficient of $\alpha \approx 3~\mu\mathrm{W}/\mathrm{K}$.

Like described before, a device with a half-circular Au layer has practical advantages over a circular Au layer. 
Hence, translating our measurement results performed in the
vacuum dewar for the device with the circular shaped Au layer to a device having only a half-cicular Au layer, 
this device should have $r_{C} = 12~\mu$m and a larger thickness $d_{Au} = 200~\mathrm{nm}$ ($V = 44~\mu\mathrm{m}^3$) 
in order to reach $T_{e} \approx T_{ph}$ with the same $\alpha \approx 3~\mu\mathrm{W}/\mathrm{K}$.  
\section{\label{sec:04}Conclusion}
In our previous measurements, we found no clear evidence of a geometrical influence of the Au layer 
on the relaxation of nonequilibrium quasiparticles \cite{Selig2014}. This is due to the fact that we performed our 
measurements in the linear regime, shown in Fig.~\ref{fig:02} for small volumes $V$. 
Hence our conclusion was that the relaxation of nonequilibrium quasiparticles does not depend on the geometry but only
on the volume of the Au layer. While this is true for the linear regime of Eq.~(\ref{eq:03}), in this work we find that 
this conclusion does not hold in the nonlinear regime. 
For the device in a vacuum dewar, in order to relax $T_{e}$ towards $T_{ph}$ we find that for a circular Au layer a volume of only
$V = 19~\mu\mathrm{m}^3$ is needed while for a half-circular Au layer an approximately twice as large 
volume of $V = 44~\mu\mathrm{m}^3$ is needed. Hence, we have determined the optimum dimensions of an either 
circular or half-circular Au energy relaxation layer to maintain a bath temperature 
quasiparticle distribution in the Nb junction when it is connected to a full superconducting NbTiN embedding 
circuit.

In the next development step we will use our findings to fabricate SIS mixer 
devices for 800~GHz to show if the nonequilibrium quasiparticle distribution in the junction can be eliminated 
by the Au energy relaxation layer in the same way when the device is irradiated with a strong RF signal. If this is 
successful we will continue this development towards a SIS mixer for 1.1 THz.
\section*{Acknowledgments}
This work is carried out within the Collaborative Research Council 956, subproject D3, funded by the Deutsche Forschungsgemeinschaft (DFG) and by BMBF, Verbundforschung Astronomie under contractno. 05A08PK2. 
The devices were fabricated in the KOSMA microfabrication laboratory and measured at the I. Physikalisches Institut, 
Universit\"at zu K\"oln. We thank R. Bruker from the Institut f\"ur Physikalische Chemie of the Universit\"at zu K\"oln for the SEM images.
\ifCLASSOPTIONcaptionsoff 
\newpage
\fi

\end{document}